\documentclass[a4paper,12pt]{article}
\usepackage{caption}
\usepackage{multicol}
\usepackage{graphicx}
\usepackage[left=3.74cm, right=2.54cm, bottom=2.54cm, top=4.54cm]{geometry}
\usepackage{amsmath}
\usepackage{setspace}
\usepackage{array}
\usepackage{subfigure}

\begin{document}
\begin{center}
{\bf \Large { Transport Coefficients in Quasi-particle Models} }
\vspace{10mm}

{ Waseem Bashir,\, Saeed Uddin{\footnote{$\it saeed\textunderscore jmi@yahoo.co.in$}}}\\
\vspace{2mm}
{ Department of Physics}\\
Jamia Millia Islamia, New Delhi-110025\\
 India
\end{center}
\vspace{2mm}
\begin{center}
{ Jan Shabir Ahmad}\\
\vspace{2mm}
Department of Physics\\
Baramulla Government Degree College\\
Baramulla, J\&K\\
India
\end{center}
\vspace{7mm}
\doublespacing
\begin{abstract}
\vspace{7mm}
\noindent We study the non-equilibrium properties of a dynamical fluid composed of quasi-particles whose mass is temperature and charge chemical potential dependent, in kinetic theory under the relaxation time approximation. In particular we calculate the scaling behaviour of bulk viscosity `$\zeta$' near the QCD chiral phase transition in the 3d $\it O$(2) universality class. It is found that the bulk viscosity `$\zeta$' does not show a divergent behaviour near the QCD chiral phase transition. This scaling behaviour of `$\zeta$' prevails  in the presence of Gold-stone modes that arise due to the explicit breaking of continuous $\it O$(4) symmetry. On contrary these modes have a significant effect on the scaling behaviour of specific heat $C_V$ which diverges at the critical temperature $T_C$.                        
\end{abstract}
\newpage
\noindent{\large \bf Introduction}\\

\vspace{3mm}
 \noindent Transport properties of a strongly interacting matter, encoded in coefficients such as shear and bulk viscosities, which describe the hydrodynamical response of the system to energy and momentum density fluctuations are of particular importance for understanding the nature of QCD matter. Their values and properties not only carry information on how far the system appears from an ideal hydrodynamics but can also provide a relevant insight into the fluid dynamics and its critical phenomena \cite{songheinz2008, son2005, hirano2006}. Even though tremendous efforts have been put forward from both theoretical and experimental sides, their exact determination still remains a challenging task, especially in the regime of experimentally accessible temperature and densities.

\noindent Recently in an analysis of the non-equilibrium properties of a quasi-particle medium in the kinetic theory, under the relaxation time approximation, the scaling behaviour of bulk viscosity $\zeta$ near the QCD chiral phase transition was studied \cite{sasaki2009}. Using the static critical exponents for 3d ${\it O}(4)$ symmetric spin models, it was pointed out that the bulk viscosity does not show the singular behaviour near the QCD chiral phase transition, in contradiction to the results obtained using the Staggered and Wilson fermion formulation of lattice QCD  \cite {karsch2008}\cite{meyer2008}, wherein a sharp rise of $\zeta$ at T = T$_C$ was reported. It was assumed that an extension of relaxation time approximation that allows for the proper incorporation of long range fluctuations of soft modes might result in the divergences of $\zeta$ at the second order ${\it O}$(4) transition. 

\noindent In general the QCD phase transition in the limit of $n_f$ massless quark flavors is controlled by $SU_{L}(n_{f}) \times SU_{R}(n_{f})$ chiral symmetry. Renormalization group arguments suggest that QCD with three degenerate light quark flavors has a first order phase transition \cite{pisarski1984}, whereas the two flavor theory is expected to have a second order phase transition. In the latter case the $SU_{L}(2) \times SU_{R}(2)$ chiral symmetry is isomorphic to ${\it O}(4)$ and the transition therefore is expected to belong to the same universality class as three dimensional ${\it O}(4)$ symmetric spin models \cite{wilczek1992}\cite{rajagopal1993}. Depending on the value of strange quark mass the QCD phase transition in the limit of vanishing light quark masses may be first order or continuous transition still belonging to three-dimensional, ${\it O}(4)$ universality class \cite{pisarski1984}. 

\noindent However none of the above mentioned lattice discretization schemes (Wilson/ Staggered) for the fermion sector of QCD preserve the full chiral symmetry of QCD Lagrangian. At non-zero lattice spacing the Wilson fermion formulation does not preserve any continuous symmetry related to the chiral sector of the QCD, while as the Staggered fermion formulation preserves at least an ${\it O}(2)$ symmetry, that gives rise to the massless Gold-stone modes in the chiral limit. Due to the existence of these modes singularities are expected on the whole coexistence line, for the ${\it O}(N)$ models with N $>$ 1 and dimensions 2 $<$ d $\leq$ 4  \cite{justin1996} \cite{basu1999}, in addition to the known singular behaviour at T$_C$. For N=2 and d=3 there exists a rigorous proof \cite{penrose1975}\cite{dunlop1975}.

\noindent Keeping into account the above mentioned facts, 1. presence of only ${\it O}(2) $ symmetry in case of staggered fermion formulation of lattice QCD,   2. the generation of Gold-stone modes due to spontaneous breaking of continuous ${\it O}(4)$ symmetry to ${\it O}(2)$ and 3. the resulting additional singularities due to these modes, we therefore calculate the scaling behaviour of bulk viscosity $\zeta$ for a quasi-particle medium in the 3d ${\it O}(2)$ universality class and evaluate the possible role of the massless Gold-stone modes on the scaling behaviour of $\zeta$ near the QCD chiral phase transition.\\
                                                                                                              
\noindent {\large \bf I. \hspace{5mm} Shear and  bulk viscosities in a quasi-particle model.}
\vspace{3mm}

\noindent  Let us consider a dynamical fluid composed of quasi-particles where the mass of each quasi-particle is temperature and charge chemical potential dependent. The calculation of the transport coefficients bulk viscosity $\zeta$ and the coefficient of viscosity $ \eta$ requires the knowledge of molecular distribution function $f ({\it p}, x )$ in terms of  which a complete description of the macroscopic state of a system is assumed to be possible.  Under the relaxation time approximation the distribution function $f({\it p}, x)$ satisfies the linear differential equation  \cite{reif1965}\\
\begin{align*}
Df \equiv \frac{\partial f}{\partial t} + {\bf v} \cdot  \frac{\partial f}{ \partial {\bf r}}+ \frac{\bf F}{m}\cdot \frac{\partial f}{\partial {\bf v}} = -\frac{f-f_0}{\tau_0} = - \frac{\delta f}{\tau_0} \tag{1}
\end{align*}

\noindent This equation characterizes the deviation of a system from the state of equilibrium, quantified by the equilibrium distribution function $f_0$. Here { m} and {\bf v} are the mass and velocity of constituents of the medium, that is under the effect of an external force {\bf F}. Here $\tau_0$ is the mean time between successive collisions among the constituents of the medium and it is by these collisions a system is assumed to restore its equilibrium state. 

\noindent For a system that is slightly removed from equilibrium, one can approximate the non-equilibrium distribution function $f ({\it p}, x)$, by the perturbative  expression
\begin{align*}
f({\it p},x)= f_0 + f_1          \tag{2}
\end{align*}

\noindent where, $f_1 << f_0$. Under this approximation the left side of Eq.1 can be approximated by neglecting terms in $f_1$ which results in an expression
\begin{align*}
Df \equiv \frac{\partial {f_0}}{\partial t} + {\bf v} \cdot  \frac{\partial {f_0}}{ \partial {\bf r}}+ \frac{\bf F}{m}\cdot \frac{\partial {f_0}}{\partial {\bf v}} =  -\frac{\delta f}{\tau _0}  \tag{3}
\end{align*}   

\noindent Here we intend to study a system which is not under the influence of any external field, which for-example in our case could be an externally applied magnetic field, therefore we can use, ${\bf F} = 0$ in Eq.(3) and we get

\begin{align*}
p^{\mu}\partial_{\mu} {f_0} = -E \frac{\delta f}{\tau_0} \tag{4}
\end{align*}

\noindent Here,  $p^{\mu} = (E, {\bf p})$ is the momentum four vector and $\partial^{\mu}= (\partial^0, \partial^i )$ is the four operator. Also we have used the relation $E /  {\bf p} = $1/{\bf v} between the energy $E= \sqrt {{\bf p}^2 + m^2}$ and momentum $\bf p$ of a relativistic particle which in absence of external force $\bf F$ is having a pure thermal motion. One can use the above calculated variation $\delta f$, to calculate the variation of energy-momentum tensor from its equilibrium value This variation of distribution function $\delta f$ can be directly used to calculate the variation of energy-momentum tensor from its equilibrium value, $\delta T^{\mu \nu}$, that in itself contains the information regarding the transport coefficients of the medium.

\noindent For a system composed of fermions and/or bosons that is slightly removed from equilibrium, the variation $\delta T^{\mu\nu}$ can be written as 

\begin{align*}
\delta T^{\mu v}= \int d \Gamma \frac{p^{\mu}p^{v}}{E}[ \delta f   +  \delta \bar f  ] \tag{5}  
\end{align*}   

\noindent where  d$\Gamma = g d^3 {\bf p} /(2 \pi)^3$ is the integration measure in the momentum space with the degeneracy factor g associated with the particle quantum numbers. If we use the value of $\delta f$ from Eq.(4) in Eq.(5) one has

\begin{align*}
\delta T^{\mu v}=- \int d \Gamma \frac{p^{\mu}p^{v}}{E^2} p^{\alpha}\partial _{\alpha} [\tau_0 f_0  + \bar \tau_0 \bar f_0 ] \tag{6}
\end{align*}  

\noindent Here $f_0$ and $\bar f_0$ are the equilibrium distribution functions for particles and anti-particles respectively and are given by   
\begin{align*}
f_0(\bar f_0)\,\,=\,\,\left [e^{{(E- {\bf p.u} \mp \mu)}/{T}} \pm 1 \right]^{-1} \tag{7}  
\end{align*}
\noindent where {\bf u} is the flow velocity and ${\bf \mu}$ is the chemical potential related to any conserved charges. The $\pm 1$ correspond to fermion and boson statistics, whereas ${\pm \mu}$ to particle and anti-particle contributions, respectively.
\noindent If one assumes that the particles and anti-particles instead of carrying a fixed value of the mass M, have a dynamical mass given by $M=M (T, \mu)$, than the above variation $\delta T^{\mu \nu}$ can be further written as \cite{sasaki2009}
\begin{align*}
\delta T ^{\mu v} = \,\,\delta T^{\mu v}_f \,\,+\,\,  \delta T^{\mu v}_{\bar f}
                  =&\,\, \int d\Gamma\,\, \frac{p^{\mu}p^{v}}{T E}f_0 (1  \pm f_0)\,\,q_f ({\bf p}, T, \mu)\,\,\\&+  \int d\Gamma\,\, \frac{p^{\mu}p^{v}}{T E}{\bar f_0} (1  \pm {\bar f_0})\,\,q_{\bar f} ({\bf p}, T, \mu)  \tag{8}
\end{align*}
\noindent where
\begin{align*}
q_{f, \bar f}({\bf p}, T, \mu) = &\left [ -\frac{{\bf p}^2}{3E} + \left ( \frac{\partial { P}}{\partial \epsilon}\right)_n \left ( E- T \frac{\partial E}{\partial T}- \mu \frac{\partial E}{\partial \mu} \right)- \left( \frac{\partial P}{\partial n}\right)_{\epsilon}\left (\frac{\partial E}{\partial \mu} \mp 1 \right) \right ]{\partial_k} {u_l} \delta ^{kl} \\&- \frac{p_k p_l}{2E} W^{kl}
\end{align*}
\noindent and 
\begin{align*}
W_{kl} = \partial_ku^l + \partial_l u^k - \frac{2}{3} \delta_{kl} \partial_i u^i   
\end{align*}

\noindent  using in Eq.(8) the energy conservation $\int d\Gamma E \left[\tau_0 f_0 \left(1 \pm f_0 \right)\,q_f  + {\bar \tau_0} {\bar f_0} \left (1 \pm {\bar f_0} \right)\,q_{\bar f}  \right] = 0 $\,\,  and comparing the resulting expression with
\begin{align*}
 \delta T^{ij}= -\zeta \delta_{ij} \partial_k u^k - \eta W_{ij}     \tag{9}
\end{align*}
\noindent where $\zeta$ and $\eta$ are the bulk and shear viscosity of the medium, one obtains following expressions for the transport coefficients of the medium \cite{sasaki2009}\\
\begin{align*}
\eta= \frac{1}{15 T} \int \frac{d^3p}{(2 \pi)^3} \frac{{\bf p}^4}{E^2} \left[g \tau f_0 (1 \pm f_0)\,\, + \,\,{\bar g} {\bar \tau_0} {\bar f_0}(1 \pm {\bar f_0})\right]   \tag{10}
\end{align*}
\noindent and
\begin{align*}
\zeta =&-\frac{1}{3T} \int \frac{d^3p}{(2 \pi )^3}   \frac{M^2}{E} [g \tau_0 f_0 (1 \pm f_0) + {\bar g} {\bar \tau_0}{\bar f_0}(1 \pm f_0)] \left [\frac{{\bf p}^2}{3E} \right.\\&\left.- \left(\frac{\partial P}{\partial \epsilon} \right)_n  \left (E - T \frac{\partial E}{\partial T}- \mu \frac{\partial E}{\partial \mu} \right) + \left(\frac{\partial P}{\partial n} \right)_{\epsilon} \frac{\partial E}{\partial \mu}\right] \\&-\frac{M^2}{E} [g \tau f_0 (1 \pm f_0) - {\bar g} {\bar \tau_0} {\bar f_0}(1 \pm {\bar f_0})] \left (\frac{\partial P}{\partial n} \right)_{\epsilon}  \tag{11}
\end{align*}

\noindent The derivatives\,\, $ \partial P/ \partial \epsilon|n$\,\, and $\partial P/ \partial n |_{\epsilon}$ which appear in Eq.(11) can be written in terms of various susceptibilities \,\,$\chi_{xy}= \partial ^2 P/ \partial x \partial y$ \,\, as follows 
\begin{align*}
\left (\frac{\partial P}{\partial \epsilon} \right)_n =\,\,\, \frac{s \chi_{\mu \mu}- n \chi_{\mu T}}{C_{V} \chi_{\mu \mu}}  \tag{12}
\end{align*}

\noindent and
\begin{align*}
\left( \frac{\partial P}{\partial n}\right)_{\epsilon}= \,\,\,\frac{n T \chi_{TT} +\,\, (n \mu  -sT)\chi_{\mu T}- s \mu \chi_{\mu \mu}}{C_V \chi_{\mu \mu}}   \tag{13}
\end{align*}

\noindent Here\,\, $n= {\partial P}/{\partial \mu}$ \,\, is the net particle number density and $s= {\partial P}/ {\partial T}$ is the entropy density. The specifiv heat $C_V$ is given by 
\begin{align*}
C_V= T \left [\chi_{TT}- \frac{\chi ^2_{\mu T}}{\chi _{\mu \mu}}   \right]   \tag{14}
\end{align*}
\noindent using the expressions of Eq.(12)- Eq.(14) we will now study the scaling behaviour of transport coefficients near the quark- hadron phase transition region.\\\\\\ 
\noindent {\large \bf II.\hspace{5mm}  Transport coefficients near the QCD chiral phase transition. }\\

\vspace{2mm}
\noindent To calculate the transport coefficients for a thermodynamic system,  one requires to work with a specific model for particle interactions. The effect of these interactions is to change some of the observables of the system for example, the particle mass M and charge chemical potential $\mu$ \cite{walecka1986}\cite{was2012}, which serve as an input in the determination of equilibrium distribution functions \,\,$f_0 , {\bar f_0},$\,\,  of particles and anti-particles respectively. These distribution functions together with the thermodynamic variables, for example temperature T and energy E can be used to determine the transport coefficients \,$\zeta$\, and \,$\eta$ for a given medium. However there are some generic properties of bulk viscosity \, $\zeta$\, that can be studied in a model independent way, through the universality arguments. In this context, of particular interest is behaviour of \, $\zeta$\, near the  phase transition.\\\\\\
\noindent{\large \bf III. \hspace{5mm}  Scaling analysis of bulk viscosity $\zeta$. }\\

\vspace{2mm}
\noindent To study the scaling of bulk viscosity $\zeta$ we proceed as follows. First we calculate the scaling behaviour of $\zeta $ in the 3d ${\it O} (2)$ universality class without incorporating the symmetry breaking effects that are applicable in view of breakdown of continuous $ {\it O} (4)$ symmetry to ${\it O}(2)$. Then we will incorporate the effect of Gold-stone modes in our calculation and evaluate the effect of these modes on the scaling behaviour of bulk viscosity near the quark-hadron phase transition region.\\\\\\
\noindent {\large \bf IV.\hspace{5mm}   Scaling of the bulk viscosity $\zeta$ in the 3d ${\it O}(2)$ universality class.}\\ 

\vspace{2mm}
\noindent In the vicinity of the chiral phase transition, the thermodynamic Free energy receives contribution from the analytic or regular $G_r$ and non-analytic or singular part $G_s$ \cite{ejiri2006}. This non-analytic part $G_s$ of free energy  results in the singular behaviour of different observables near the phase transition \cite{gottlieb1987} \cite{gott1988} \cite{gavai1989}\cite{ejiri2006}. To parametrize the scaling behaviour of $G_s$ near the chiral phase transition let us consider a general thermodynamic scaling formula for the singular part of free energy  \cite{peskinschroder} 
\begin{align*}
G_s (M,t) = t^{\beta (1 + \delta)} {\hat f} (M t^{-\beta})             \tag{15}
\end{align*}
Here $\beta$ and $\delta$ are the critical exponents for a given universality class. M  is the order parameter of the transition and \, `t' \, is the reduced temperature variable defined as $t= \bar t + A \,\bar\mu^2$ \cite {ejiri2006} where , $\bar t= (T-T_c)$/$T_C$ and $\bar \mu= \mu/ T_C$.  In the discussion to follow we will calculate all the relevant quantities required for examining the scaling behaviour of bulk viscosity $\zeta$, using the scaling form of $G_s$ as mentioned above. We will limit our discussion to the $\mu =0$ case only.

\noindent Using the relation $H= \partial G_s/{\partial M}$ , where magnetic field H is a variable conjugate to M,\, together with scaling law for $G_s$ as given by Eq.(15) we obtain
\begin{align*}
H= t^{\beta \delta} {\hat f'} \left(M t^{-\beta}  \right)  \tag{16}
\end{align*}
Here ${\hat f'}$ is partial derivative of $f$ with respect to M. Now calculating the inverse of Eq.(16) we get the scaling law for M
\begin{align*}
M= t^{\beta} {\hat c} \left(H t^{- \beta \delta} \right)  \tag{17}
\end{align*}
where $\hat c = \hat f^{-1}$.
Now we will use the scaling function $G_s$ as defined in Eq.(15) together with generallised reduced temperature  $t = \bar t + A \bar \mu^2$ and calculate all the relevant quantities required for examining the scaling behaviour of bulk viscosity $\zeta$. In the following we will consider only the transition point at $\mu =$ 0. As already mentioned we are dealing with a system which is not under the influence of any external force therefore we will use H=0.   

\noindent For T $\rightarrow$ T$_C$ at zero field H, the scaling behaviour of order parameter M is
\begin{align*}
M \sim t^{\beta}\, \hat c (0) \sim t^{\beta}  \tag{19}
\end{align*} 

\noindent The scaling behaviour of ${\partial M}$/${\partial T}$ is
\begin{align*}
\frac{\partial M}{\partial T} \sim t^{\beta -1}  \tag{20}
\end{align*}  

\noindent Using the scaling behaviour of M and ${\partial M}$/ ${\partial T}$, the scaling of susceptibility $\chi _{\mu \mu} \simeq \partial ^2 G_s / \partial \mu ^2$ can be readily computed.

\noindent Taking double derivative of $G_s$ with respect to $\mu$ we get 
\begin{align*}
\chi_{\mu \mu}\simeq\,\,\,\, &t^{[\beta(1+\delta)-2]} \left[f(M t^{-\beta})+ 2\hat f(M t^{-\beta})t^{-\beta}\right]\left({\partial t}/{\partial \mu}\right)^2 \\&+t^{[\beta(1+\delta)-1]} \left[\hat f(Mt^{-\beta})+ \hat f'(M t^{-\beta}) t^{-\beta} \right] \left({\partial^2 t}/{\partial \mu}^2\right)\\& + t^{[\beta \delta -1]} \left[\hat f'' (Mt^{-\beta}) \right] \left({\partial t}/{\partial \mu}\right) 
\end{align*} 
 

\noindent Now since $t\, =\, \bar t + A \bar \mu^2\,= \,\,\,( T- T_c )/T_c \,+\, A\, (\mu/ T_c) ^2$, we have  $\partial t/ \partial \mu= 2A \mu/ T_c $ and $\partial^2t/ \partial \mu^2 = 2A/T_c$. Using these expressions together with the scaling of M as in Eq. (19) the scaling form of $\chi_{\mu \mu}$ becomes
\begin{align*}
\chi_{\mu \mu} \sim t^{1- \alpha}  \tag{21}
\end{align*} 
where we have used the relation $\beta (1+ \delta)= 2- \alpha$. Next we calculate the scaling form of $\chi_{TT} \simeq \partial^2 G_s / \partial T^2$. Using the scaling function $G_s$ as in Eq.(15) and taking double derivative with respect to chemical potential $\mu$ we get.
\begin{align*}
\chi_{TT} \,\,\simeq\,\,\,\, &  t^{[\beta(1+\delta)-2]} \left[\hat f(Mt^{-\beta}) +2 \hat f'(Mt^{-\beta})t^{-\beta}+ t^{-2 \beta} \hat f''(M t^{-\beta}) \right]\left({\partial t}/{\partial T}\right)^2 \\&+ 
t^{[\beta (1 + \delta)-1]} \left[\hat f(Mt^{-\beta}) + t^{-\beta} \hat f(M t^{-\beta})\right] \left({\partial ^2 t}/{\partial T^2}\right)\\&
+t^{[\beta \delta -1]} \left[ \hat f'(M t^{-\beta}) \right] \left({\partial M}/{\partial T}\right)\left({\partial t}/{\partial T} \right)
\end{align*}

\noindent Using ${\partial t}/{\partial \mu}= 2A \mu /T_c$, together with the scaling law for order parameter M as in Eq.(19) we get
\begin{align*}
\chi_{TT} \simeq t^{-\alpha}  \tag{22}
\end{align*} 
Now we calculate the scaling of susceptibility $\chi_{\mu T}= {\partial^2 G_s}/{\partial \mu \partial T}$. Using again the scaling function of $G_s$ as in Eq.(15) we get.
\begin{align*}
\chi_{\mu T} \simeq & \,\,t^{[\beta(1+ \delta)-1]}\left[ \hat f (M t^{-\beta})+ 2 \hat f(M t^{-\beta})t^{-\beta}+ \hat f''(M t^{-\beta}) t^{-2 \beta}\right]\left({\partial t}/{ \partial \mu}\right) \left({\partial t}/{\partial T} \right)\\&+ \,\, t^{[\beta(1+\delta)-1]} \left[\hat f(M t^{-\beta}) + \hat f' (M t^{-\beta})t^{-\beta} \right] {\partial ^2 t}/{\partial \mu \partial T}
\end{align*}

\noindent After substituting the values of different derivatives  we get
\begin{align*}
\chi_{\mu T} \simeq 0  \tag{23}
\end{align*}

\noindent Therefore the scaling behaviour of specific heat $C_V$ given by Eq.(14) near the phase transition at $\mu$=0 becomes
\begin{align*}
C_V= T \left [ \chi_{TT}- \frac{\chi_{\mu T}^2}{\chi_{\mu \mu}}  \right] \simeq t^{-\alpha}  \tag{24}
\end{align*} 

\noindent Using the scaling laws for $\chi_{\mu \mu}$, $\chi_{\mu T}$, $\chi_{T T}$ and specific heat $C_V$, the scaling behaviour of bulk viscosity $\zeta$ is readily calculable. From Eq.(11) the divergent behaviour of $\zeta$ can be written as \cite{sasaki2009}

\begin{align*}
\zeta^{div} \sim -\frac{M^2}{E} \left[-\left(\frac{\partial P}{\partial \varepsilon}\right)_n \left(E- T \frac{\partial E}{\partial T}-\mu \frac{\partial E}{\partial \mu} \right)+ \left(\frac{\partial P}{\partial n}\right)_\varepsilon \frac{\partial E}{\partial \mu}- \left(\frac{\partial P}{\partial n}\right)_{\varepsilon}\right]  \tag{25}
\end{align*}

\noindent Using Eq.(12), Eq.(13) together with the scaling laws of $\chi_{\mu \mu}$, $\chi_{\mu T}$, $\chi_{T T}$ and $C_V$ we have
\begin{align*}
\left( \frac{\partial P}{\partial \varepsilon}\right)_n= \frac{s\chi_{\mu \mu}- n\chi_{\mu T}}{C_V \chi_{\mu \mu}} \sim t^{\alpha}\\\\
\left(\frac{\partial P}{\partial n} \right)_{\varepsilon}= \frac{nT \chi_{TT}+ (n \mu - s T)\chi_{\mu T}- s \mu\chi_{\mu \mu}}{C_V \chi_{\mu \mu}} = 0
\end{align*}
\vspace{3mm}

\noindent Also we have the scaling relations,\, ${\partial E}/{\partial T}= ({M}/{E}) ({\partial M}/{\partial T}) \sim t^{2 \beta -1} $ and\,\, ${\partial E}/{\partial \mu}= ({M}/{E}) ({\partial M}/{\partial \mu}) \sim t^{2 \beta -1} $. 
Therefore the scaling of bulk viscosity $\zeta$ is given by the relation
\begin{align*}
\zeta^{div} \sim t^{4 \beta + \alpha -1}    \tag{26}
\end{align*}

\noindent Now using the critical exponents for 3d {{\it O}(2)} model, $\alpha$= -0.01722 and $\beta$= 0.349, the singular part of bulk viscosity, $\zeta ^{div}$, vanishes near the phase transition region.
\begin{table}[h]
\caption { \bf Critical exponents $\alpha$, $\beta$, $\gamma$, $\delta$ and the universal constant $\tilde c_2$ for three-dimensional ${\it O}(4)$ and ${\it O}(2)$ universality class  \cite{engels2003} \cite{engels2001}. The critical exponents are connected by the relations $\gamma= \beta (\delta -1)$ and $ \beta (1 + \delta)= 2 - \alpha$. }
\vspace{14mm}
\centering
\begin{tabular}{c c c c c c}
\hline\hline
Model\,\,\,\,& $\alpha$\,\,\,\, & $\beta$\,\,\,\, & $\gamma$\,\,\,\, & $\delta$\,\,\,\, & $\tilde c_2$\\
\hline
{\it O}(4)\,\,\,\, & -0.21\,\,\,\, & 0.380\,\,\,\, & 1.453 \,\,\,\,& 4.824 \,\,\,\,& 0.666(6)  \\

{\it O}(2)\,\,\,\, & -0.01722\,\,\,\, & 0.349\,\,\,\, & 1.319 \,\,\,\,& 4.780 \,\,\,\,& 0.592(10)  \\
\hline
\end{tabular}
\end{table}\\

\vspace{5mm}
\noindent {\large \bf V. \hspace{5mm}  Effect of Gold-stone modes on the scaling of bulk viscosity $\zeta$}.\\ 

\vspace{3mm}
\noindent The spontaneous breaking of a continuous {\it O}(N) symmetry results in the generation of Gold-stone modes. For an ${\it O}(N)$ invariant nonlinear $\sigma$-models these modes lead to the singular behaviour of susceptibilities near the phase co-existence line \cite{engmen2000} \cite{engelholt2000}. To look for any singular behaviour of bulk viscosity $\zeta$ at the QCD phase transition, that  might arise due to the contribution from these modes, we consider a general Widom -Griffith's form of equation of state, which describes the critical behaviour of magnetization in the vicinity of T$_C$ and is compatible with the singularities due to these Gold-stone modes.

\noindent It is given by \cite{griffiths1967}
\begin{align*}
y = f(x)  \tag{27}
\end{align*}   
where the variables y and x are given by
\begin{align*}
y \equiv h/ M^{\delta}, \hspace{15mm} x \equiv t/ M^{1/ {\beta}}
\end{align*}
The variables t and h are the normalized reduced temperature $t=(T- T_C)/ T_0$ and the magnetic field $h= H/H_0$. The normalization  of function $ f$ are taken to be
\begin{align*}
f(0)=1, \hspace{15mm} f(-1)= 0 \end{align*}
From the critical exponents $\beta$ and $\delta$ one can derive all other critical exponents. The various critical exponents are related as
\begin{align*}
d \nu= \beta (1 + \delta), \hspace{15mm} \gamma = \beta (\delta -1), \hspace{15mm} \nu_c =\nu/ {\beta \delta}
\end{align*}

\noindent Another way of expressing the dependence of magnetization M on the reduced temperature variable t and magnetic field h is 
\begin{align*}
M = h^{1/\delta} f_G (t/ h^{1/{\beta \delta}}) \tag{28}
\end{align*}
where, $f_G $ is the scaling function. The scaling forms in Eq.(27) and Eq.(28) are clearly equivalent since the variables x and y are related to the scaling function $f_G$ and its argument by
\begin{align*}
y = f_G^{-\delta}, \hspace{15mm} x= (t/ h^{1/ {\beta \delta}}) f_G^ {-1/{\beta}}
\end{align*}
In presence of Gold-stone modes the scaling function $f_G$ is given by \cite{ejirikl2009}
\begin{align*}
f_G(z) = (-z)^{\beta} (1+ \tilde c_2 \beta (-z)^{{-\beta \delta}/2}) \tag{29}
\end{align*}
where $z= t/h^{1/{\beta \delta}}$.
As has been discussed in \cite{engels2001}, Eq.(29) can be easily obtained from the magnetic equation of state derived by Wallace and Zia \cite{wallace1975}. The universal amplitude $\tilde c_2$ is given in Table 1.

\noindent Integrating the expression $M \sim \partial G_s/ \partial H$, where magnetization M and scaling function $f_G$ are defined by Eq.(28) and Eq.(29) respectively, the free energy G$_s$, comes out to be
\begin{align*}
G_s= h(-t)^{\beta} + 2/3\,\, \tilde c_2 \,\beta\, (-t)^{\beta - \beta \delta /2}\, h^{3/2} \tag{30}
\end{align*}

\noindent To describe the QCD phase transition which is controlled by two independent parameters, charge chemical potential $\mu$ and temperature T, we once again chose to parametrize the reduced temperature variable t as, t= $\bar t$ + A\,$\bar \mu^2$, with $\bar t$= $(T-T_C)$/ T$_C$
and $\bar \mu=$ $\mu$/$T_C$. Using the above form of free energy G$_s$, we will calculate all the relevant quantities required for examining the scaling behaviour of $\zeta$. As in previous case we will again limit our study to $\mu=0$ case only.

\noindent First we calculate the scaling of susceptibility  $\chi_{TT} \simeq \partial ^2 G_s/ \partial  T^2$. Evaluating the double derivative of G$_s$ with respect to temperature T , we get
\begin{align*}
\chi_{T T} \simeq &\,\, (-t)^{[\beta -2]} \left[ 1 + \tilde c_2 (-t)^{{-\beta \delta}/2}  \right] \left( - {\partial t}/{ \partial T} \right)^2 \\&+ (-t)^ {[\beta -1]} \left [ 1 + \tilde c_2 (-t)^{{-\beta \delta }/2} \right]
\left( {- \partial ^2 t}/{\partial T^2}\right)
\end{align*}

\noindent Using the first and second order derivatives of the reduced temperature variable t, ${\partial t}/{\partial T}$= T$_C^{-1}$ and $\partial^2t $/${\partial T^2}$= 0 respectively, the scaling form of susceptibility $\chi_{T T}$ becomes 
\begin{align*}
\chi_{TT} \sim (-t)^{\beta -2} \tag{31}
\end{align*}
  
\noindent Next we calculate the scaling behaviour of susceptibility $\chi_{\mu \mu} \simeq \partial ^2 G_s/ \partial \mu ^2$. Defining a double derivative of G$_s$ with respect to baryon chemical potential $\mu$, we get

\begin{align*}
\chi_{\mu \mu}\, \simeq\,\, & (-t)^{[\beta-2]} \left[1 + \tilde c_2 (-t)^{{\beta \delta}/2}\right] \left( -{\partial t}/{\partial \mu}\right)^2 \\&
+ (-t)^{[\beta-1]} \left[1+\tilde c_2(-t)^{{-\beta \delta}/2} \right]\left({\partial ^2 t}/{\partial \mu^2} \right)
\end{align*}

\noindent Using the relations $\partial t/ \partial \mu = 2A\mu / T_C$ = 0 and $\partial ^2 t/ \partial \mu^2= 2A/T_C$, the scaling form of susceptibility $\chi_{\mu \mu}$ becomes
\begin{align*}
\chi_{\mu \mu} \sim (-t) ^{\beta -1} \tag{32}
\end{align*}

\noindent Now finally we calculate the scaling behaviour of $\chi_{\mu T}\simeq \partial ^2 G_s/ \partial \mu \partial T$. Evaluating the double derivative of free energy G$_s$ with respect to $\mu$ and T we get.
\begin{align*}
\chi_{\mu T} \,\simeq &\,\, (-t)^{[\beta-2]} \left[1 + \tilde c_2 (-t)^{{-\beta \delta }/2}\right] \left( {\partial t}/{\partial \mu}\right)\left({\partial t}/{\partial T} \right)\\& + (-t)^{[\beta-1]} \left[1+ \tilde c_2 (-t)^{{-\beta \delta}/2}\right] \left(-{\partial ^2 t}/{\partial \mu \partial T} \right)
\end{align*}

\noindent Using the relations $\partial t/ \partial T= T_C^{-1}$, \, $\partial ^2 t/ \partial \mu \partial T= 0$ and $\partial t/ \partial \mu= 2 A \mu T_C^{-1}$ we get
\begin{align*}
\chi_{\mu T} = 0 \tag{33}
\end{align*}
Therefore the scaling behaviour of specific heat $C_V$ defined in Eq.(14) is given by 
\begin{align*}
C_V = T\left[\chi_{TT}- \frac{\chi_{\mu T}^2}{\chi_{\mu \mu}} \right] \sim (-t)^{\beta -2} \tag{34}
\end{align*} 
which for the static critical exponents of 3d $\it O$(2) model, has a divergent behaviour near the QCD phase transition. Now as already mentioned in previous section  the divergent behaviour of bulk viscosity $\zeta$ at critical point is controlled by the term\\

\begin{align*}
\zeta^{div} \sim -\frac{M^2}{E} \left[ - \left(\frac{\partial P}{\partial \varepsilon} \right)_n \left(E - T \frac{\partial E}{\partial T} - \mu \frac{\partial E}{\partial \mu} \right)+ \left(\frac{\partial P}{ \partial n} \right)_{\varepsilon} \frac{\partial E}{\partial \mu} - \left(\frac{\partial P}{\partial n} \right)_{\varepsilon} \right] \tag{35}
\end{align*}\\ 
Using again Eq.(12) and Eq.(13) together with the scaling laws for susceptibilities $\chi_{\mu \mu}, \chi_{\mu T}, \chi_{T T}$ and specific heat $C_V$ we have
\begin{align*}
\left(\frac{\partial P}{\partial \varepsilon} \right)_n = \frac{s \chi_{\mu \mu}- n \chi_{\mu T}}{C_V \chi_{\mu \mu}} \sim (-t)^{\beta -2}
\end{align*}
\begin{align*}
\left(\frac{\partial P}{\partial n} \right)_{\varepsilon}= \frac{nT \chi_{TT} + (n \mu - sT)\chi_{\mu T}- s \mu \chi_{\mu \mu}}{C_V \chi_{\mu \mu}} = 0
\end{align*}
Also, using Eq.(28) in conjugation with Eq.(29) we have the scaling relations ${\partial E}/ {\partial T}= (M/E) ({\partial M}/{\partial T}) \sim (-t)^{2\beta -1}$ and ${\partial E}/{\partial \mu}= ({M}/{E})({\partial M}/{\partial \mu}) \sim (-t)^{2 \beta -1}$.
Therefore the singular part of the bulk viscosity $\zeta$ scales as
\begin{align*}
\zeta^{div} \sim (-t)^{3 \beta +1} \tag{36}
\end{align*}
Now using the static critical exponents for 3d {\it O}(2) model, given in Table 1, the singular part of the bulk viscosity, $\zeta^{div}$, vanishes near the QCD phase transition.\\\\\\
{\large \bf VI.\hspace{5mm}  Results and Discussions.}\\

\vspace{2mm}
\noindent We studied the non-equilibrium properties of a quasi-particle medium at finite temperature and density in a kinetic theory under the relaxation time approximation, where the mass of each quasi-particle is assumed to be temperature and chemical potential dependent. In particular we studied the scaling behaviour of bulk viscosity $\zeta$ near the QCD chiral phase transition in the 3d ${\it O}(2)$ universality class. It is found that the bulk viscosity $\zeta$ approaches to a finite value, $\zeta^{regular}= \zeta-\left(\zeta^{div}=0\right)$, near the QCD chiral phase transition. This scaling  behaviour of $\zeta$ prevails even after incorporating the effect of Gold-stone modes, that arise due to the explicit breaking of the continuous ${\it O}(4)$ symmetry in the staggered fermion formulation of the lattice QCD at non-zero lattice spacing. A similar non-divergent result for $\zeta$ is reported in ref \cite{gubser2008}, where it is found that the the ratio $\zeta$/s (here s is the entropy density) has a maximum value of \,\,$\zeta/s|_{max} \simeq$ 0.06\,\, at critical temperature T$_C$, in contradiction to the results reported in ref \cite{sasaki2009}, wherein a sharp rise of fraction $\zeta$/s near the QCD phase transition was obtained, using an analysis based on the exact sum rule for QCD with 2 + 1 flavors, lattice QCD data on $\varepsilon -3P$ and a simple model for the spectral function. This non-divergent behaviour of $\zeta$ at critical temperature T$_C$ using the non-perturbative method of Ads/CFT correspondence is similar to leading order perturbative QCD result of $\zeta$/s $\simeq$ 0.02 $\alpha_s^2$ \cite{arnold2006} for a coupling strength of  \,\,0.06 $ < \alpha_s <  $ 0.3.
On the other hand the Gold-stone modes are found to have a significant effect on the scaling behaviour of the specific heat `$C_V$' near the QCD phase transition. These modes lead to the divergent behaviour of the specific heat at the critical temperature $T_C$. However these divergences do not appear in the bulk viscosity as the scaling of `$\zeta$' in our quasi-particle model is always determined by the product $\zeta^{div} \sim \left({M^3}/{C_V}\right)\left ({\partial M}/{\partial T} + {\partial M}/{\partial \mu}\right)$,  consequently $\zeta^{div} \propto C_V^{-1}$, which is different than the scaling behaviour of  `$\zeta$' given by $\zeta^{div} \propto C_V$, as found in ref \cite{karsch2008}. This behaviour of bulk viscosity under the kinetic theory in relaxation time approximation should not be regarded as essentially new. In ref \cite{weinberg1971} it is shown that a gas of structure less point particles have negligible bulk viscosity both in the non-relativistic and extreme relativistic limits. It is particularly due to this reason that mostly bulk viscosity $\zeta$ is often neglected even when the shear viscosity is taken into account. However it should be mentioned that our analysis of scaling behaviour of bulk viscosity $\zeta$ is valid only under the relaxation time approximation. An extension beyond this approximation that allows for a proper incorporation of the long range fluctuations of the soft modes might possibly result in the divergent behaviour of bulk viscosity $\zeta$ at the QCD phase transition, with the divergent behaviour of bulk viscosity being controlled by the dynamic critical exponents rather than the static critical exponents used in this study. \\\\\\

\noindent {\Large \bf   Acknowledgments}\\

\vspace{3mm}
\noindent  Waseem Bashir is thankful to the University Grants Commision for providing Project Fellowship, Saeed Uddin is thankful to the University Grants Commision (UGC), New Delhi, for the Major Research Project grant. Jan Shabir Ahmad is greatful to University Grants Commision, New Delhi, for the financial assistance during the period of deputation.  

\bibliographystyle{unsrt}
\bibliography{transport}

\end{document}